\documentclass[journal,10pt]{IEEEtran}
\pdfoutput=1

\usepackage{etex}				

\usepackage{times}
\usepackage{graphicx}
\usepackage{graphics}
\usepackage{epsfig}
\usepackage{amssymb}
\usepackage{amsmath}
\usepackage{amsthm}
\usepackage{color}
\usepackage{fancyhdr}
\usepackage[colorlinks,urlcolor=blue] {hyperref}
\usepackage{url}
\usepackage{amsfonts}		
\usepackage{mathrsfs}
\usepackage{textcomp}
\usepackage{latexsym}		
\usepackage{amsthm}
\usepackage{wrapfig}
\usepackage{pdfpages}
\usepackage{caption}
\usepackage{subfig}

\usepackage{tikz}
\usetikzlibrary{chains}
\usetikzlibrary{fit}
\usepackage{pgflibraryarrows}		
\usepackage{pgflibrarysnakes}		
\usepackage{xcolor}
\usepackage{epsfig}

\usepackage{listings}
\usepackage{url}

\usepackage[latin1]{inputenc}

\usepackage{paralist} 

\usepackage{mathabx} 

\def\Hb{\ensuremath{H_{\rm b}}}
\def\S{\mathsf{S}}
\def\D{\mathsf{D}}
\def\C{\mathsf{C}}

\def\E{\mathsf{E}}
\def\F{\mathsf{F}}
\def\R{\mathsf{R}}

\def\Rset{\mathcal{R}_\epsilon}
\def\Dset{\mathcal{D}_\epsilon}
\def\Bparam{Bhattacharyya parameter}

\newcommand{\supp}{\textnormal{supp}}
\newcommand{\W}{\mathsf{W}} 
\newcommand{\Hh}[1]{H\!\left({#1}\right)} 
\newcommand{\Hcond}[2]{H\!\left({#1}|{#2}\right)} 
\newcommand{\Prvcond}[2]{\,{\rm Pr}\!\left[\left.#1\,\right|\,#2\right]} 
\newcommand{\Prv}[1]{\,{\rm Pr}\!\left[#1\right]} 
\newcommand{\SNR}{\textnormal{SNR}}
\newcommand{\Bernoulli}[1]{\textnormal{Bernoulli}\left(#1\right)}  
\newcommand{\pos}[2]{\textnormal{pos}_{{#1}}\left({#2}\right)} 

\begin{document}

\newtheorem{mythm}{Theorem}
\newtheorem{myprop}{Proposition}
\newtheorem{mycor}{Corollary}
\newtheorem{mylem}{Lemma}
\newtheorem{myclaim}{Claim}
\newtheorem{mysubclaim}{Subclaim}
\newtheorem{myfact}{Fact}
\newtheorem{myconj}{Conjecture}

\theoremstyle{definition}
\newtheorem{mydef}{Definition}
\newtheorem{myex}{Example}
%
\title{Achieving the Capacity of any DMC using only Polar Codes}

\author{David~Sutter,
  ~Joseph~M.~Renes,
  ~Fr\'ed\'eric~Dupuis,

  ~and~Renato~Renner
  \thanks{Part of this paper will be presented at the IEEE
    Information Theory Workshop (ITW), Lausanne, Switzerland,
    2012. This work was supported by the Swiss National Science
    Foundation (SNF) through the National Centre of Competence in
    Research ``Quantum Science and Technology'' and project
    No. 200020-135048 and PP00P2-128455, by the European Research
    Council (ERC) via grant No. 258932, and by the German Science
    Foundation (grants No. \mbox{CH~843/1-1} and \mbox{CH~843/2-1}).}
  \thanks{D.\ Sutter is with the Department of Information Technology
    and Electrical Engineering, ETH Zurich, Switzerland (e-mail:
    suttedav@student.ethz.ch).}
  \thanks{J.M.\ Renes, F.\ Dupuis and R.\ Renner are with the
    Institute of Theoretical Physics, ETH Zurich, Switzerland (e-mail:
    renes@itp.phys.ethz.ch; dupuis@phys.ethz.ch;
    renner@phys.ethz.ch).}}



\maketitle

\begin{abstract}
We construct a channel coding scheme to achieve the capacity of any discrete memoryless channel based solely on the techniques of polar coding. In particular, we show how source polarization and randomness extraction via polarization can be employed to ``shape'' uniformly-distributed i.i.d.\ random variables into approximate i.i.d.\ random variables distributed according to the capacity-achieving distribution. We then combine this shaper with a variant of polar channel coding, constructed by the duality with source coding, to achieve the channel capacity. Our scheme inherits the low complexity encoder and decoder of polar coding. It differs conceptually from Gallager's method for achieving capacity, and we discuss the advantages and disadvantages of the two schemes. An application to the AWGN channel is discussed.
\end{abstract}


\IEEEpeerreviewmaketitle

\section{Introduction} \label{sec: intro}
\IEEEPARstart{P}{olar} codes, introduced by Ar{\i}kan \cite{arikan09}, are the first set of codes that provably achieve the symmetric capacity\footnote{The symmetric capacity of a DMC is the mutual information of the channel output given a \emph{uniform} input.} of any discrete memoryless channel (DMC)~\cite{sasoglu2009}, using encoding and decoding algorithms whose complexity is essentially linear in the blocklength $N$.\footnote{The precise encoding and decoding complexity is $O(N\log N)$.} By now, the polarization phenomenon at the heart of polar coding has been adapted for use in a variety of information-processing tasks. 

Being a family of linear codes, polar codes do not achieve the true channel capacity whenever the optimum input distribution is not uniform, which is generically the case for arbitrary DMCs. As noted in~\cite{sasoglu2009}, Gallager's method~\cite[p.208]{gallager68} of ``shaping'' blocks of independent  uniformly-distributed encoded message bits into (a rational approximation to) an arbitrary distribution of a channel input symbol can be combined with polar coding to approach the channel capacity. The shaper essentially creates a super-channel whose optimal input distribution is uniform, so  that concatenating the usual multi-bit polar encoder with the shaper results in an encoder suitable for approaching capacity. The overhead of the shaper complicates the encoding and decoding algorithms, though does not affect the scaling of the complexity in the blocklength for fixed accuracy in approximating the non-uniform distribution. 

Here we use the techniques of polar coding to give a more information-theoretic shaper construction and exhibit a modified family of polar codes which can achieve the capacity of any DMC. Instead of approximating a single input-bit, our shaper approximates a string of i.i.d. input-bits. Compared to Gallager's method, this leads to a  conceptually different coding scheme having better encoding and decoding complexity. (See Section~\ref{sec: DiffGallager} for a comparison of the methods.) 

 The idea of our shaper is to run a randomness extractor for the optimal input distribution in reverse, a technique previously exploited by two of us to construct capacity-achieving codes in the context of one-shot channel coding~\cite{renes11}. As in~\cite{renes11}, we construct the outer polar code\footnote{The outer polar code is the code for the super-channel.} by exploiting the duality between channel coding and source coding with side information, detailed for polar coding in~\cite{arikan10}.

To understand the main idea more concretely, suppose that $\W:\mathcal{X}\to \mathcal{Y}$ denotes a DMC with binary input alphabet $\mathcal{X}=\{0,1\}$, arbitrary output alphabet $\mathcal{Y}$ and transition probabilities $\W(y|x), x\in \mathcal{X}, y\in \mathcal{Y}$. $\W^L$ denotes the channel corresponding to $L$ uses of $\W$. We consider binary-input DMCs only for convenience; the techniques of \cite{sasoglu2009} and \cite{karzand10} can be used to generalize the scheme to DMCs with arbitrary input size. Furthermore, let $\Bernoulli{p}$ for $p\in \left[0,1\right]$ be the capacity-achieving input distribution, so that $I(X{:}Y)=C(\W)$, for $X\sim\Bernoulli{p}$ and $Y=\W(X)$. Given $L$ i.i.d.\ instances of $X$, roughly $H(X^L)=L\Hb(p)$ approximately-uniformly distributed bits can be extracted, where $\Hb$ denotes the binary entropy~\cite{elias1972}. 
Heuristically, we may thus hope to simulate $X^L$ by inputting $L\Hb(p)$ \emph{uniform} bits to the inverse of the extractor. 

Given $X^L$, an extractor function may be stochastically run in
reverse by making use of the joint distribution of its inputs and
outputs.  Given an extractor output value, an input value is chosen
randomly among the preimages according to the conditional distribution
induced from the joint distribution by fixing the output
value. However, it is not clear this process can be done efficiently
for arbitrary input distributions.

Luckily, this process is efficient for extractors based on the source
polarization phenomenon. A polarization extractor for $X^L$ simply
generates $U^L=X^L G_L$ (when $L=2^\ell$ for $\ell\in\mathbb{Z}^+$)
using the channel transform $G_L=\bigl( \begin{smallmatrix} 1 &0\\
  1&1\end{smallmatrix} \bigr)^{\otimes \, \ell} $ and keeps only those
$U_i$ such that $\Hcond{U_i}{U^{i-1}} \geq 1-\epsilon$ for some
specified $\epsilon$. Polarization ensures that there will be roughly
$L\Hb(p)$ such $U_i$.\footnote{Note that this is not a randomness
  extractor in the usual sense, which is designed to work for
  \emph{any} input distribution of sufficiently high
  min-entropy~\cite{shaltiel2004}.}  To invert this extractor, we
first build up a vector $\hat{U}^L$ by filling with
uniformly-distributed input the positions $i$ for which $U_i|U^{i-1}$
has entropy at least $1-\epsilon$ and stochastically generating the
remaining positions using the distributions of the $U_i|U^{i-1}$. The
output $\hat{X}^L$ is just $\hat{X}^L=\hat{U}^LG_L$, and, for $\epsilon$
small, closely approximates $X^L$.\footnote{Korada and Urbanke apply a similar construction, which they called \emph{randomized rounding}, to the problem of lossy source coding in~\cite{korada_polar_2010}.} The necessary distributions can be
efficiently computed, a feature used in the similarly-constructed
decompressor of polar source coding~\cite{arikan10}.

Combining the shaper with the channel $\W^L$ creates a super-channel
$\W'_{K,L}$, to which the usual polar coding techniques could be
applied. However, this does not result in an efficient coding scheme
because the likelihoods and {\Bparam}s of $\W'$ are not necessarily
easy to compute. To regain efficiency, we instead employ a polar
coding scheme adapted from the source compression scheme for $U^L$
given $Y^L$ at the decompressor. Due to its i.i.d.\ structure, the
necessary parameters can be efficiently computed, meaning that the
complexity of the resulting decoder will again be essentially linear
in the number of uses of the channel $\W$.

This paper is structured as follows. In Section~\ref{sec:shaping} we define the shaper and super-channel precisely.  
Section~\ref{sec:scheme} details our coding scheme, Section~\ref{sec: schemeworks} shows that it achieves the capacity of any binary-input DMC, and Section~\ref{sec: encodingDecoding} shows that it is reliable. Section~\ref{sec:efficiency} then describes how encoding, decoding, and channel construction can be performed efficiently. Section~\ref{sec:derand} demonstrates that the shaper can be almost completely derandomized without impacting the code performance. Section~\ref{sec: DiffGallager} explains the differences between the new scheme and Gallager's method. Finally in Section~\ref{sec: discussion} we discuss some possible modifications of the new scheme as well as some potential applications, in particular communication over the AWGN channel with an average power constraint.

\section{Polarization-Based Shaper and Super-Channel}
\label{sec:shaping}

We briefly recount the use of source polarization in randomness extraction~\cite{arikan10, abbe11, abbe11_2} and then formulate the shaper and super-channel. First it is convenient to introduce the following notation. Let $[k]=\left \lbrace 1,\ldots,k \right \rbrace$. For $x \in \mathbb{F}_2^k$ and $\mathcal{I}\subseteq [k] $ we have $x[\mathcal{I}]=[x_i:i\in \mathcal{I}]$ and $x^i=[x_1,\ldots,x_i]$.
For an ordered set of distinct elements $\mathcal{A} \subseteq [k]$ and $a\in \mathcal{A}$, $\pos{\mathcal{A}}{a}$ denotes the position of the entry $a$ in  $\mathcal{A}$. 

As described above, a $K$-bit polarization extractor $\E_{L,K}$ for
$X^L$ simply outputs the $K$ bits of $U^L=X^LG_L$ for which
$H(U_i|U^{i-1})$ are greatest. We denote this (ordered) set of indices
by $\mathcal{E}_K$ and the output of the extractor by
$U^L[\mathcal{E}_K]$. 

\begin{figure}[!htb]
\centering
\tikzstyle{block1} = [draw, rectangle, 
    minimum height=2em, minimum width=3em, node distance=2.5cm,anchor=center]
\def\gap{1}

\begin{tikzpicture}[auto,>=latex']
	
    \node (input) at (-3.5*\gap,0) {};
    \node [block1] (recon) at (-1*\gap,0) {$G_L$};
    \node [block1] (polar) at (1*\gap,0) {$\F_{L,K}$};
    \node (output) at (3.8*\gap,0) {};

    \draw [->,] (input) -- node {{$X^L$}} (recon); 
    \draw [->,] (recon) -- node {${U}^L$} (polar);
    \draw [->,] (polar) -- node {${U}^L[\mathcal{E}_K]$} (output);
   
    \node[draw,dashed,fit=(recon) (polar),inner sep=8pt] (shaper) {};
    \node[below of=shaper] {$\E_{L,K}$};
    
\end{tikzpicture}
\caption{\small Polarization-based randomness extractor $\E_{L,K}$. The input $X^L$ is first transformed to $U^L$ via the polarization transformation $G_L$, and subsequently $\F_{L,K}$ filters out the $K$ bits of $U^L$ for which $H(U_i|U^{i-1})$ are greatest.}
\label{fig: Extraction}
\end{figure}
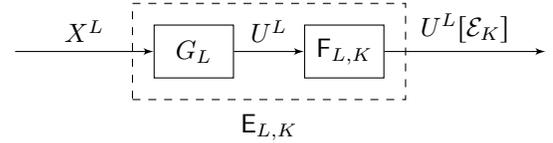
 The aim of randomness extraction is to output $K$ approximately uniform bits, where the approximation is quantified using the variational distance. Recall that for distributions $P$ and $Q$ over the same alphabet $\mathcal{X}$, the variational distance is defined by $\delta(P,Q):=\frac12\sum_{x\in\mathcal{X}}\left|P(x)-Q(x) \right| $. We will often abuse notation slightly and write a random variable instead of its distribution in $\delta$. 
 
Using $\mathcal{E}_K$ we define the shaper for $X^L$ as follows
\begin{mydef}
\label{def: shaper}
The shaper  $\S_{K,L}$ for $X^L$ is the map $\S_{K,L}:\mathcal{U}^K\to\mathcal{X}^L$ taking input $U^K$ to  $\hat{X}^L=\hat{U}^LG_L$, with 
\begin{align}
\label{eq:shaperdef}
\hat{U}_i=\left\{
\begin{array}{ll} U_{{\rm pos}_{\mathcal{E}_K}(i)}& i\in\mathcal{E}_K\\ Z_i & {\rm else}
\end{array}\right..
\end{align}
Here $Z_i$ is a random variable generated from the distribution of $U_i|U^{i-1}$, using $U^L=X^LG_L$. 
\end{mydef}

\begin{figure}[!htb]
\centering
\tikzstyle{block1} = [draw, rectangle, 
    minimum height=2em, minimum width=3em, node distance=2.5cm,anchor=center]
\def\gap{1}

\begin{tikzpicture}[auto,>=latex']
	
    \node (input) at (-3.8*\gap,0) {};
    \node [block1] (recon) at (-1*\gap,0) {$\R_{K,L}$};
    \node [block1] (polar) at (1*\gap,0) {$G_L$};
    \node (output) at (3.5*\gap,0) {};

    \draw [->] (input) -- node {{${\tilde{U}}^K$}} (recon); 
    \draw [->] (recon) -- node {$\hat{U}^L$} (polar);
    \draw [->] (polar) -- node {$\hat{X}^L$} (output);
   
    \node[draw,dashed,fit=(recon) (polar),inner sep=8pt] (shaper) {};
    \node[below of=shaper] {$\S_{K,L}$};
   
\end{tikzpicture}
\caption{\small Generation of an approximation to $X^L$ from a uniform input $\tilde{U}^K$ using the shaper $\S_{K,L}$. $\hat{U}^L$ is first constructed by $\R_{K,L}$ from the uniform input according to (\ref{eq:shaperdef}). Applying $G_L$ gives $\hat{X}^L$, which has nearly the same distribution as $X^L$. }
\label{fig: ExactRec}
\end{figure}
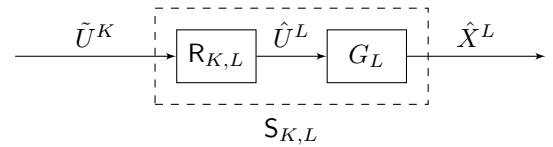

Using the shaper with uniform input $\tilde{U}^K$ (a $K$-bit vector
whose entries are i.i.d. $\Bernoulli{\frac12}$) generates an
approximation $\hat{X}^L:=\S_{K,L}(\tilde{U}^K)$ to $X^L$ (see also~\cite[Lemma 11]{korada_polar_2010}).

\begin{mylem}
  \label{lem:shapergood} For $\epsilon \geq 0$ and $K$ such that
  $\Hcond{U_i}{U^{i-1}} \geq 1-\epsilon$ for all $i \in
  \mathcal{E}_K$,
  \begin{align*}
 \delta\left(\hat{X}^L,X^L\right)\leq K
  \sqrt{\frac{\ln 2}{2} \epsilon} \ .
  \end{align*}
\end{mylem}

\begin{IEEEproof}
  Let $\hat{U}^L$ be the $L$-bit string obtained when using the shaper
  with uniform input $\tilde{U}^K$ (cf.\ Eq.~\ref{eq:shaperdef}).  We
  have $X^L = U^L G_L$ and $\hat{X}^L = \hat{U}^L G_L$ and, hence,
\begin{align}
  \delta(\hat{X}^L, X^L) =  \delta(\hat{U}^L, U^L) \ .
\end{align}
We will bound the distance on the right hand side. For this, we
introduce a family of intermediate distributions $P^{(i)}_{U_1 \cdots
  U_i \hat{U}_{i+1} \cdots \hat{U}_L}$, for $i=0, \ldots, N$, defined
by
\begin{align}
  P^{(i)}_{U_1 \cdots U_i \hat{U}_{i+1} \cdots \hat{U}_L} := P_{U_1
    \cdots U_i } P_{\hat{U}_{i+1} \cdots \hat{U}_L | \hat{U}_1 \cdots
    \hat{U}_i}\ ,
\end{align}
so that $P^{(0)}_{\hat{U}_{1} \cdots \hat{U}_L} = P_{\hat{U}_1 \cdots
  \hat{U}_L}$ and $P^{(L)}_{U_1 \cdots U_L} = P_{U_1 \cdots U_L}$.
 By the triangle inequality,
\begin{align}
  \delta(\hat{U}^L, U^L) 
& \leq
  \sum_{i = 1}^{L}  \delta(P^{(i-1)}_{U_1 \cdots U_{i-1} \hat{U}_{i}
    \cdots \hat{U}_L}, P^{(i)}_{U_1 \cdots U_{i} \hat{U}_{i+1}
    \cdots \hat{U}_L}) \\
  & \leq \sum_{i = 1}^{L}  \delta(P^{(i-1)}_{U_1 \cdots U_{i-1}
    \hat{U}_{i}}, P^{(i)}_{U_1 \cdots U_{i-1} U_{i} }) \ , \label{eq_Usum}
\end{align}
where the last line follows from the fact that the variational
distance is non-increasing under stochastic maps~\cite{liesevajda87}
(we apply this to the map that generates $\hat{U}_{i+1} \cdots
\hat{U}_L$ according to the distribution $P_{\hat{U}_{i+1} \cdots
  \hat{U}_L| \hat{U}_{1} \cdots \hat{U}_{i}}$).  Each term of the sum
can be written as $ \delta(P_{U^{i-1}}  P_{\hat{U}_i|\hat{U}^{i-1}}
     , P_{U^{i-1}}  P_{U_i|U^{i-1}})$ or, equivalently, $\mathsf{E}_{U^{i-1}} \left[\delta(P_{\hat{U}_i|\hat{U}^{i-1}}, P_{U_i|U^{i-1} })\right]$. 
     To bound this, we use
Pinsker's inequality~\cite[p.58]{csiszarkorner81} as well as the concavity of the square root,
\begin{align}
  &\mathsf{E}_{U^{i-1}}\left[ \delta(P_{\hat{U}_i|\hat{U}^{i-1}}, P_{U_i|U^{i-1} })  \right] \nonumber \\
  &\hspace{10mm}\leq \mathsf{E}_{U^{i-1}}\left[\sqrt{\tfrac{\ln
      2}{2}D(P_{U_i|U^{i-1}}\|P_{\hat{U}_i|\hat{U}^{i-1}})}\right]\\
  &\hspace{10mm}\leq \sqrt{\tfrac{\ln
      2}{2}\mathsf{E}_{U^{i-1}}\left[
    D(P_{U_i|U^{i-1}}\|P_{\hat{U}_i|\hat{U}^{i-1}})\right]
    } \ .
\end{align}
By construction, the conditional
distribution of $\hat{U}_i$ for all $i \in \mathcal{E}_K$ is the uniform
distribution, so that
\begin{align}
 \mathsf{E}_{U^{i-1}}\left[
    D(P_{U_i|U^{i-1}}\|P_{\hat{U}_i|\hat{U}^{i-1}})
 \right]
   &= 1 - H(U_i|U^{i-1}) \\ 
   &\leq \epsilon \ .   
\end{align}
Furthermore, for all $i \notin \mathcal{E}_K$, the conditional
distribution of $\hat{U}_i$ equals $P_{U_i|U^{i-1}}$, so that the
corresponding term in the sum~\eqref{eq_Usum} vanishes. The sum can
thus be rewritten as
\begin{align}
  \delta(\hat{U}^L, U^L)
\leq 
  \sum_{i \in \mathcal{E}_K} \sqrt{\tfrac{\ln 2}{2}
    \epsilon} \ ,
\end{align}
from which the assertion follows. 
\end{IEEEproof}

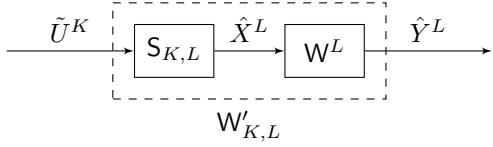
\begin{figure}[!htb]
	\centering
\tikzstyle{block1} = [draw, rectangle, 
    minimum height=2em, minimum width=3em, node distance=2.5cm,anchor=center]
\def\gap{1}

\begin{tikzpicture}[auto,>=latex']
	
    \node (input) at (-3.35*\gap,0) {};
    \node [block1] (shaper) at (-1*\gap,0) {$\S_{K,L}$};
    \node [block1] (channel) at (1*\gap,0) {$\W^L$};
    \node (output) at (3.35*\gap,0) {};

    \draw [->] (input) -- node {{$\tilde{U}^K$}} (shaper); 
    \draw [->] (shaper) -- node {$\hat X^L$} (channel);
    \draw [->] (channel) -- node {$\hat{Y}^L$} (output);
   
    \node[draw,dashed,fit=(shaper) (channel),inner sep=8pt] (super) {};
    \node[below of=super] {$\W'_{K,L}$};
\end{tikzpicture}
        \caption{\small The super-channel $\W'_{K,L}:=\W^L\circ\S_{K,L}$, shown here acting on the uniformly-random input $\tilde{U}^K$, which results in $\hat{Y}^L$.}
        \label{fig: super-channel}
\end{figure}

Concatenating the shaper with the channel gives the super-channel
$\W'_{K,L}:=\W^L \circ\S_{K,L}$. Monotonicity of the variational
distance gives the following lemma, which is the basis of our coding
scheme. Letting $\hat{Y}^L:=\W^L(\hat{X}^L)$ and $Y^L =\W^L(X^L)$, we
have
\begin{mylem} 
\label{lem:inputoutput}
For $\epsilon \geq 0$ and $K$ such that
  $\Hcond{U_i}{U^{i-1}} \geq 1-\epsilon$ for all $i \in
  \mathcal{E}_K$,
\begin{align*}
\delta\!\left((\tilde{U}^K,\hat{Y}^L),(U^L[\mathcal{E}_K],Y^L)\right)\leq
K\sqrt{\frac{\ln 2}{2} \epsilon} \ .
\end{align*}
\end{mylem}
\begin{IEEEproof}
  Lemma~\ref{lem:shapergood} implies
  $\delta((\hat{X}^L,\hat{Y}^L),(X^L,Y^L))\leq \epsilon'$ by the
  monotonicity of the variational distance under stochastic
  maps. Applying $G_L$ to $X^L$ or $\hat{X}^L$ and marginalizing over
  the elements not in $\mathcal{E}_K$ is also a stochastic map, so
  $\delta((\hat{U}^L[\mathcal{E}_K],\hat{Y}^L),(U^L[\mathcal{E}_K],Y^L))\leq
  \epsilon'$. Observing that $\hat{U}^L[\mathcal{E}_K]=\tilde{U}^K$
  completes the proof.
\end{IEEEproof}

\section{Coding Scheme}
\label{sec:scheme}
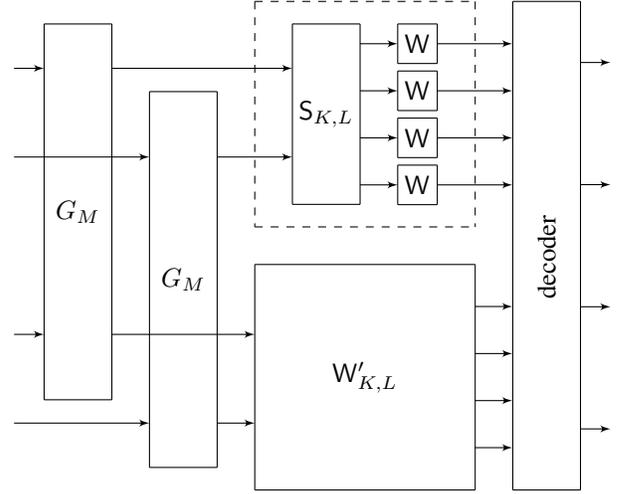
\begin{figure}[!t]
	\centering
	\def \xcomp{0.9}
\def \ycomp{5}
\def \xblock{2.93}
\def \yblock{3}

\def \xgapd{0.5} 
\def \ygapd{0.9} 

\def \ybs{-0.3} 

\def \ys{0.1} 

\def \xdec{0.9} 
\def \w{0.526} 

\def \xs{0.4} 

\begin{tikzpicture}[scale=1,auto, node distance=1cm,>=latex']
	
     \draw [draw] (0,0) -- (\xcomp,0); 
     \draw [draw] (0,0) -- (0,-\ycomp);
     \draw [draw] (0,-\ycomp) -- (\xcomp,-\ycomp);
     \draw [draw] (\xcomp,0) -- (\xcomp,-\ycomp);     
    \node at (0.5*\xcomp,-0.5*\ycomp) {$G_M$};
    
     \draw [draw] (\xgapd+\xcomp,-\ygapd) -- (2*\xcomp+\xgapd,-\ygapd); 
     \draw [draw] (\xgapd+\xcomp,-\ygapd) -- (\xgapd+\xcomp,-\ycomp-\ygapd);
     \draw [draw] (\xgapd+\xcomp,-\ycomp-\ygapd) -- (2*\xcomp+\xgapd,-\ycomp-\ygapd);
     \draw [draw] (2*\xcomp+\xgapd,-\ygapd) -- (2*\xcomp+\xgapd,-\ycomp-\ygapd);     
    \node at (1.5*\xcomp+\xgapd,-0.5*\ycomp-\ygapd) {$G_M$};
   
      \draw [draw,dashed] (2*\xcomp+2*\xgapd,-\ybs) -- (2*\xcomp+2*\xgapd+\xblock,-\ybs);   
      \draw [draw,dashed] (2*\xcomp+2*\xgapd,-\ybs-\yblock) -- (2*\xcomp+2*\xgapd+\xblock,-\ybs-\yblock);   
      \draw [draw,dashed] (2*\xcomp+2*\xgapd,-\ybs) -- (2*\xcomp+2*\xgapd,-\ybs-\yblock);   
      \draw [draw,dashed] (2*\xcomp+2*\xgapd+\xblock,-\ybs) -- (2*\xcomp+2*\xgapd+\xblock,-\ybs-\yblock);   
      
      \draw [draw] (2*\xcomp+2*\xgapd,\ybs-\ycomp-\ygapd) -- (2*\xcomp+2*\xgapd+\xblock,\ybs-\ycomp-\ygapd);         
      \draw [draw] (2*\xcomp+2*\xgapd,\ybs-\ycomp-\ygapd+\yblock) -- (2*\xcomp+2*\xgapd+\xblock,\ybs-\ycomp-\ygapd+\yblock);   
      \draw [draw] (2*\xcomp+2*\xgapd,\ybs-\ycomp-\ygapd) -- (2*\xcomp+2*\xgapd,\ybs-\ycomp-\ygapd+\yblock);
      \draw [draw] (2*\xcomp+2*\xgapd+\xblock,\ybs-\ycomp-\ygapd) -- (2*\xcomp+2*\xgapd+\xblock,\ybs-\ycomp-\ygapd+\yblock);           
         \node at (2*\xcomp+2*\xgapd+0.5*\xblock,\ybs-\ycomp-\ygapd+0.5*\yblock) {$\W'_{K,L}$};    

   \draw [draw] (2*\xcomp+3*\xgapd+\xblock,-\ybs) -- (2*\xcomp+3*\xgapd+\xblock+\xdec,-\ybs);   
   \draw [draw] (2*\xcomp+3*\xgapd+\xblock,-\ygapd-\ycomp+\ybs) -- (2*\xcomp+3*\xgapd+\xblock+\xdec,-\ygapd-\ycomp+\ybs);    
   \draw [draw] (2*\xcomp+3*\xgapd+\xblock,-\ybs) -- (2*\xcomp+3*\xgapd+\xblock,-\ygapd-\ycomp+\ybs);  
   \draw [draw] (2*\xcomp+3*\xgapd+\xblock+\xdec,-\ygapd-\ycomp+\ybs) -- (2*\xcomp+3*\xgapd+\xblock+\xdec,-\ybs);   
   \node [rotate=90] at (2*\xcomp+3*\xgapd+\xblock+0.5*\xdec,-0.5*\ygapd-0.5*\ycomp+0.5*\ybs) {decoder};

  \draw [draw] (2*\xcomp+3*\xgapd,0) -- (2*\xcomp+3*\xgapd+\xcomp,0);  
  \draw [draw] (2*\xcomp+3*\xgapd,-\yblock-2*\ybs) -- (2*\xcomp+3*\xgapd+\xcomp,-\yblock-2*\ybs);
  \draw [draw] (2*\xcomp+3*\xgapd,0) -- (2*\xcomp+3*\xgapd,-\yblock-2*\ybs);  
 \draw [draw] (2*\xcomp+3*\xgapd+\xcomp,-\yblock-2*\ybs) -- (2*\xcomp+3*\xgapd+\xcomp,0);
     \node [] at (2*\xcomp+3*\xgapd+0.5*\xcomp,-0.5*\yblock-\ybs) {$\S_{K,L}$};
     
    \draw [draw] (3*\xcomp+4*\xgapd,0) -- (3*\xcomp+4*\xgapd+\w,0);  
    \draw [draw] (3*\xcomp+4*\xgapd,-\w) -- (3*\xcomp+4*\xgapd+\w,-\w);
    \draw [draw] (3*\xcomp+4*\xgapd,0) -- (3*\xcomp+4*\xgapd,-\w);
    \draw [draw] (3*\xcomp+4*\xgapd+\w,0) -- (3*\xcomp+4*\xgapd+\w,-\w);
   \node at (3*\xcomp+4*\xgapd+0.5*\w,-0.5*\w) {$\W$};
   
    \draw [draw] (3*\xcomp+4*\xgapd,-\ys-\w) -- (3*\xcomp+4*\xgapd+\w,-\ys-\w);  
    \draw [draw] (3*\xcomp+4*\xgapd,-\w-\ys-\w) -- (3*\xcomp+4*\xgapd+\w,-2*\w-\ys);
    \draw [draw] (3*\xcomp+4*\xgapd,-\ys-\w) -- (3*\xcomp+4*\xgapd,-2*\w-\ys);
    \draw [draw] (3*\xcomp+4*\xgapd+\w,-\ys-\w) -- (3*\xcomp+4*\xgapd+\w,-2*\w-\ys);
   \node at (3*\xcomp+4*\xgapd+0.5*\w,-1.5*\w-\ys) {$\W$};
   
    \draw [draw] (3*\xcomp+4*\xgapd,-2*\ys-2*\w) -- (3*\xcomp+4*\xgapd+\w,-2*\ys-2*\w);  
    \draw [draw] (3*\xcomp+4*\xgapd,-3*\w-2*\ys) -- (3*\xcomp+4*\xgapd+\w,-3*\w-2*\ys);
    \draw [draw] (3*\xcomp+4*\xgapd,-2*\ys-2*\w) -- (3*\xcomp+4*\xgapd,-3*\w-2*\ys);
    \draw [draw] (3*\xcomp+4*\xgapd+\w,-2*\ys-2*\w) -- (3*\xcomp+4*\xgapd+\w,-3*\w-2*\ys);
   \node at (3*\xcomp+4*\xgapd+0.5*\w,-2.5*\w-2*\ys) {$\W$};
   
    \draw [draw] (3*\xcomp+4*\xgapd,-3*\ys-3*\w) -- (3*\xcomp+4*\xgapd+\w,-3*\ys-3*\w);  
    \draw [draw] (3*\xcomp+4*\xgapd,-4*\w-3*\ys) -- (3*\xcomp+4*\xgapd+\w,-4*\w-3*\ys);
    \draw [draw] (3*\xcomp+4*\xgapd,-3*\ys-3*\w) -- (3*\xcomp+4*\xgapd,-4*\w-3*\ys);
    \draw [draw] (3*\xcomp+4*\xgapd+\w,-3*\ys-3*\w) -- (3*\xcomp+4*\xgapd+\w,-4*\w-3*\ys);
   \node at (3*\xcomp+4*\xgapd+0.5*\w,-3.5*\w-3*\ys) {$\W$};

    \draw [draw,->] (-\xs,-0.1*\ycomp-0.1*\ygapd) -- (0,-0.1*\ycomp-0.1*\ygapd); 
    \draw [draw,->] (\xcomp,-0.1*\ycomp-0.1*\ygapd) -- (2*\xcomp+3*\xgapd,-0.1*\ycomp-0.1*\ygapd); 
     
    \draw [draw,->] (-\xs,-0.3*\ycomp-0.3*\ygapd) -- (\xcomp+\xgapd,-0.3*\ycomp-0.3*\ygapd);  
    \draw [draw,->] (2*\xcomp+\xgapd,-0.3*\ycomp-0.3*\ygapd) -- (2*\xcomp+3*\xgapd,-0.3*\ycomp-0.3*\ygapd); 
        
    \draw [draw,->] (-\xs,-0.7*\ycomp-0.7*\ygapd) -- (0,-0.7*\ycomp-0.7*\ygapd); 
    \draw [draw,->] (\xcomp,-0.7*\ycomp-0.7*\ygapd) -- (2*\xcomp+2*\xgapd,-0.7*\ycomp-0.7*\ygapd); 
        
    \draw [draw,->] (-\xs,-0.9*\ycomp-0.9*\ygapd) -- (\xcomp+\xgapd,-0.9*\ycomp-0.9*\ygapd);  
    \draw [draw,->] (2*\xcomp+\xgapd,-0.9*\ycomp-0.9*\ygapd) -- (2*\xcomp+2*\xgapd,-0.9*\ycomp-0.9*\ygapd);      
    
    \draw [draw,->] (3*\xcomp+3*\xgapd,-0.5*\w) -- (3*\xcomp+4*\xgapd,-0.5*\w); 
    \draw [draw,->] (3*\xcomp+4*\xgapd+\w,-0.5*\w) -- (3*\xcomp+6*\xgapd+\w,-0.5*\w); 
    
    \draw [draw,->] (3*\xcomp+3*\xgapd,-1.5*\w-\ys) -- (3*\xcomp+4*\xgapd,-1.5*\w-\ys); 
    \draw [draw,->] (3*\xcomp+4*\xgapd +\w,-1.5*\w-\ys) -- (3*\xcomp+6*\xgapd+\w,-1.5*\w-\ys);
        
    \draw [draw,->] (3*\xcomp+3*\xgapd,-2.5*\w-2*\ys) -- (3*\xcomp+4*\xgapd,-2.5*\w-2*\ys); 
    \draw [draw,->] (3*\xcomp+4*\xgapd+\w,-2.5*\w-2*\ys) -- (3*\xcomp+6*\xgapd+\w,-2.5*\w-2*\ys);
     
    \draw [draw,->] (3*\xcomp+3*\xgapd,-3.5*\w-3*\ys) -- (3*\xcomp+4*\xgapd,-3.5*\w-3*\ys);             
    \draw [draw,->] (3*\xcomp+4*\xgapd+\w,-3.5*\w-3*\ys) -- (3*\xcomp+6*\xgapd+\w,-3.5*\w-3*\ys);
    
      \draw [draw,->] (2*\xcomp+2*\xgapd+\xblock,-\ygapd-\ycomp+0.5*\w) -- (3*\xcomp+6*\xgapd+\w,-\ygapd-\ycomp+0.5*\w);
      \draw [draw,->] (2*\xcomp+2*\xgapd+\xblock,-\ygapd-\ycomp+1.5*\w+\ys) -- (3*\xcomp+6*\xgapd+\w,-\ygapd-\ycomp+1.5*\w+\ys);
      \draw [draw,->] (2*\xcomp+2*\xgapd+\xblock,-\ygapd-\ycomp+2.5*\w+2*\ys) -- (3*\xcomp+6*\xgapd+\w,-\ygapd-\ycomp+2.5*\w+2*\ys);
      \draw [draw,->] (2*\xcomp+2*\xgapd+\xblock,-\ygapd-\ycomp+3.5*\w+3*\ys) -- (3*\xcomp+6*\xgapd+\w,-\ygapd-\ycomp+3.5*\w+3*\ys);

      \draw [draw,->] (3*\xgapd+2*\xcomp+\xdec+\xblock,-\ybs+0.125*2*\ybs-0.125*\ycomp-0.125*\ygapd) -- (3*\xgapd+2*\xcomp+\xdec+\xblock+\xs,-\ybs+0.125*2*\ybs-0.125*\ycomp-0.125*\ygapd);
      \draw [draw,->] (3*\xgapd+2*\xcomp+\xdec+\xblock,-\ybs+0.375*2*\ybs-0.375*\ycomp-0.375*\ygapd) -- (3*\xgapd+2*\xcomp+\xdec+\xblock+\xs,-\ybs+0.375*2*\ybs-0.375*\ycomp-0.375*\ygapd);
      \draw [draw,->] (3*\xgapd+2*\xcomp+\xdec+\xblock,-\ybs+0.625*2*\ybs-0.625*\ycomp-0.625*\ygapd) -- (3*\xgapd+2*\xcomp+\xdec+\xblock+\xs,-\ybs+0.625*2*\ybs-0.625*\ycomp-0.625*\ygapd);
      \draw [draw,->] (3*\xgapd+2*\xcomp+\xdec+\xblock,-\ybs+0.875*2*\ybs-0.875*\ycomp-0.875*\ygapd) -- (3*\xgapd+2*\xcomp+\xdec+\xblock+\xs,-\ybs+0.875*2*\ybs-0.875*\ycomp-0.875*\ygapd);

\end{tikzpicture}
        \caption{\small The coding scheme for $L=4$, $M=2$ and
          $K=2$. At the outer layer, polar codes are used to provide
          reliable communication over the super-channel $\W'_{K,L}$,
          by using the multilevel coding method to treat it as a
          sequence of binary input channels. The encoder and decoder
          are constructed from the compressor and decompressor for the
          task of compressing $U^L[\Rset]$ relative to side
          information $Y^L$ at the decoder; in particular, the frozen
          input bits correspond to the compressor outputs. Here
          $U^L[\mathcal{E}_K]$ is the output of the polarization-based
          randomness extractor applied to the random variable $X^L$,
          which has the optimal distribution for achieving the
          capacity of the physical channel $\W$, and $Y^L$ is the
          corresponding channel output. At the inner layer,
          polarization is again used to shape the uniform inputs from
          the outer layer into a good approximation to $X^L$ for
          transmission over $\W$.}
        \label{fig: newCodingScheme}
\end{figure}

As in Gallager's original approach, our coding scheme is based on concatenating an outer coding layer for reliable transmission through the super-channel with an inner shaping layer to realize $\W'_{K,L}$. In principle, polar codes may be employed for this purpose, using the multilevel coding described in~\cite[Section III.B]{sasoglu2009}.\footnote{This type of multilevel coding is due to Imai and Hirakawa \cite{imai77}.} There, a channel with multiple input bits (assumed to be uniformly distributed) is decomposed into a sequence of binary-input channels and usual polar coding is applied to each. In the present context, the $j$th such channel ${\W'}_{\!\!K,L}^{(j)}$ maps $\tilde{U}_j$ to $(\W'_{K,L}(\tilde{U}^K),\tilde{U}^{j-1})$. Letting $M$  be the number of super-channel uses, the overall blocklength is then $N:=ML$. Figure~\ref{fig: newCodingScheme} depicts the case $M=2$, $L=4$, and $K=2$.

However, to apply the polar coding construction we would need to know both the output Bhattacharyya parameters (for code construction) and input likelihood ratios (for decoding) of each ${\W'}_{\!\!K,L}^{(j)}$. 
These might not be efficiently computable from the properties of $W$ itself, as the shaper output is not precisely $X^L$. 
Instead, we will use the close relationship between channel coding and source coding with side information~\cite{arikan10,renes11} to construct a reliable and efficient scheme. 

Consider the general problem of compressing a uniformly-distributed bit $U$ relative to arbitrary side information $Y$, where $Y=\W(U)$ for some channel $\W$.  Suppose that we have a compressor / decompressor pair $(\C,\D)$ such that $U^M$ can be reconstructed from $Y^M$ and the compressor output $\C(U^M)$ with probability $1-P_{\rm err}$, i.e. ${\rm Pr}[U^M\neq \D(Y^M,\C(U^M))]=P_{\rm err}$. Each compressor output $c$ defines a set of codewords: all the values of $u^M$ for which $\C(u^M)=c$. Choosing a compressor output at random, encoding messages into the associated codewords, and decoding them with the decompressor $\D$ then leads to a block error probability (averaged over uniformly-chosen input messages and codebooks) of $P_{\rm err}$~\cite[Lemma 2]{renes11}.\footnote{Note that transforming this code into one with small worst-case error probability would still require an expurgation argument.}

Therefore, in order to construct an efficient and reliable coding scheme for the super-channel, we look for an efficient and reliable compression scheme for $\tilde{U}^K$ relative to $\hat{Y}^L$. Due to Lemma~\ref{lem:inputoutput}, any compression scheme for  $U^L[\mathcal{E}_K]$ relative to $Y^L$ will only incur a negligible additional probability of error when applied to $(\tilde{U}^K,\hat{Y}^L)$ (cf. Theorem~\ref{thm:reliability}). Polar coding provides  such an efficient and reliable scheme. Thus, by assuming the model $(U^L[\mathcal{E}_K],Y^L)$ instead of the true parameters $(\tilde{U}^K,\hat{Y}^L)$,  the super-channel decompressor benefits from the independence of $X^L$ for efficient decompression while incurring negligible error overhead. 

To be more precise, let $V_i$ be the $i$th bit of $U^L[\mathcal{E}_{K}]$. Given $M$ copies of every $V_i$, we can use standard polar source coding on each of these sequences in turn to compress $V_i$ relative to the side-information $Y^L V^{i-1}$. The compressor outputs those bits of $T^{(i)}=V_i^MG_M$ for which $H(T_j^{(i)}|Y^MT^{j-1(i)})$ exceeds some fixed threshold $\epsilon$; call this set $\mathcal{C}_\epsilon$. The Bhattacharyya parameters and likelihood ratios associated with $(V_i,Y^LV^{i-1})$, necessary to determine $\mathcal{C}_\epsilon$ and to construct the decoder, are precisely those computed in the polar source coding scheme of $X$ relative to side information $Y$. 

To turn this into channel coding, we simply fix (freeze) the value of the bits in $\mathcal{C}_\epsilon$, use the bits in the complement $\mathcal{C}_\epsilon^c$ as data bits, and map messages to codewords by applying $G_M$. The values taken by the frozen bits are known to the decoder and one can use the source coding {decompressor} to decode the associated ${\W'}_{\!\!K,L}^{(i)}$ channel input. Note that the ${\W'}_{\!\!K,L}^{(i)}$ must be decoded in order, as $T^{(i)}$ is part of the channel output for all subsequent channels.

For each $i$ the above scheme operates at a rate of $1 - H(V_i|Y^L V^{i-1})$,  yielding a total rate per $\W'_{K,L}$ use of
\begin{align}
\label{eq:rate}
	\sum_{i=1}^K 1 - \Hcond{V_i}{Y^L V^{i-1}} = K - \Hcond{U^L[\mathcal{E}_K]}{Y^{L}}.
\end{align}
Dividing this rate by $L$ then gives the rate per use of $\W$, 
\begin{align}
R\,{:=}\lim \limits_{L\rightarrow\infty}\frac1L\left[|\mathcal{E}_K|-H(U^L[\mathcal{E}_K]|{Y}^L)\right].
\end{align}

\section{Achieving Capacity}
\label{sec: schemeworks}

We now show that a suitable choice of $K$ enables our scheme to achieve the capacity of the physical channel $\W$. 
To do so we make use of the polarization property of the $U_i|U^{i-1}$ for a given $X^L$. Consider the two (ordered) sets 
 \begin{align}
  \mathcal{R}_{\epsilon}&:=\left \lbrace i \in [L]: \, \Hcond{U_i}{U^{i-1}} \geq 1-\epsilon   \right\rbrace\quad\text{and}\\
  \mathcal{D}_{\epsilon}&:= \left \lbrace i \in [L]: \, \Hcond{U_i}{U^{i-1}} \leq \epsilon  \right\rbrace
\end{align}
of essentially random and deterministic variables, respectively. From Theorems 1 and 2 of~\cite{arikan10} we have $|\mathcal{R}_\epsilon |=L\Hb(p)-o(L)$ and $|\mathcal{D}_\epsilon|=L(1-\Hb(p))-o(L)$ with $\epsilon=O(2^{-L^\beta})$ for $\beta< \frac12$. 

As an aside, observe that choosing $\mathcal{E}_K=\Rset$ with $K=|\mathcal{R}_\epsilon|$ yields a good shaper by Lemma~\ref{lem:shapergood}, which gives the following
\begin{mythm} \label{thm:approx}
$\delta(\S_{|\mathcal{R}_\epsilon|,L}(\tilde{U}^{|\mathcal{R}_\epsilon|}),X^L)\!=\!O(L2^{-\frac12L^\beta})$ for $\beta<\frac12$.
\end{mythm}

It is simple to show that the coding scheme achieves $C(\W)$. 

\begin{mythm}
\label{thm:capacity}
$R =C(\W).$
\end{mythm}
\begin{IEEEproof}
Applying the chain rule to $H(U^L|Y^L)$ gives
\begin{align}
\Hcond{U^L}{Y^L}&=\Hcond{U^L[\Rset]}{Y^L}+\Hcond{U^L[\Rset^c]}{Y^LU^L[\Rset]} \nonumber\\
&\geq \Hcond{U^L[\Rset]}{Y^L},
\end{align}
where $\Rset^c$ is the complement of $\Rset$ in $[L]$. 
Since $H(U^L|Y^L)=H(X^L|Y^L)=LH(X|Y)$ and $H(X)=\Hb(p)$, by (\ref{eq:rate}) and the properties of $\Rset$ we find  
\begin{align}
R&\geq\lim_{L\rightarrow\infty}\frac1L\left[L\Hb(p)-o(L)-L\Hcond{X}{Y}\right]=C(\W).
\end{align}
As $R$ cannot exceed the capacity, we have $R=C(\W)$. 
\end{IEEEproof}

\section{Reliability} \label{sec: encodingDecoding}
In this section we analyze the reliability of the coding scheme, starting with a general lemma on the reliability of using the ``wrong'' compressor / decompressor pair in the problem of source coding. 
\begin{mylem}
\label{lem:wrongdecoder}
Let $X$ and ${X}'$ be arbitrary random variables such that $\delta(X',{X})\leq \eta$ and let $\W$ denote an arbitrary stochastic map. 
 If $\C$ and $\D$ are a compressor / decompressor pair for $(X,\W(X))$, such that 
${\rm Pr}[\hat{X}\neq X]\leq \eta'$ where $\hat{X}=\D(\W(X),\C(X))$, then, for $\hat{X}'= \D(\W(X'),\C(X'))$,
\begin{align*}
\Prv{\hat{X}'\neq X'}\leq \eta+\eta'.
\end{align*}
\end{mylem}
\begin{IEEEproof}
  Note that the pairs $(X, \hat{X})$ and $(X', \hat{X}')$ are obtained
  from $X$ and $X'$ by applying the stochastic map that takes $x$ to
  $(x, \mathsf{D}(\mathsf{W}(x), \mathsf{C}(x)))$. Because the
  variational distance is non-increasing under such maps, we have
  \begin{align}
    \delta((X, \hat{X}), (X', \hat{X}')) \leq \delta(X, X') \leq  \eta \ .
  \end{align}
  Furthermore, defining $(X,X)$ to be the random variable
  $(X,\bar{X})$ with distribution $P_{X \bar{X}} = P_X \delta_{X
    \bar{X}}$, we have
  \begin{align}
    \delta((X, X),  (X,\hat{X})) = \Pr[\hat{X} \neq X] \leq \eta' \ .
  \end{align}
  Hence, applying the triangle inequality, we obtain
  \begin{align}
    \delta((X, X), (X', \hat{X}')) \leq \eta + \eta' \ .
  \end{align}  
 Now note that the variational distance can also be written as
  \begin{align}
    \delta(A, A') = \sum_{a: \, P_{A}(a) \leq P_{A'}(a)} P_{A'}(a) - P_{A}(a) \ .
  \end{align}
  Applied to $A = (X, X)$ and $A'=(X', \hat{X}')$, and using that
  $P_{X X}(x, \hat{x}) = 0$ for $x \neq \hat{x}$, we immediately
  obtain
  \begin{align}
    \delta((X, X), (X', \hat{X}')) \geq \sum_{x \neq \hat{x}}
    P_{X'\hat{X}'}(x,\hat{x}) \ ,
  \end{align}
  which  implies that $\Pr[\hat{X}' \neq X'] \leq \eta + \eta'$. 
\end{IEEEproof}

Next we analyze the reliability of the multilevel coder. Suppose we
would like to compress ($L$ instances of) $(V_1,\dots,V_n)$ relative
to side information $Y$, by sequentially compressing $V_i$ relative to
$V^{i-1}Y$. Define $\hat{V}_i$ to be the output of the decompressor,
let $\mathcal{A}_i$ be the event that $\hat{V}_i \neq V_i$ (i.e. that
the decompressor makes a mistake at position $i$), and let
$\mathcal{B}_i:= \cup_{k=1}^i \mathcal{A}_k$. Note that
$\Prv{\mathcal{B}_n}$ is the probability of incorrectly decoding at
least one $V_i$ for $i \in \left[n\right]$. Let $r$ be a bound on the
probability of that we decode incorrectly at any step and that the
previous steps are all correct: $\Prv{\mathcal{A}_j \cap
  \mathcal{B}^c_{j-1}}\leq r$ for all $j\in[n]$. Then
\begin{mylem} \label{lem: newerrLem}
For $n\in \mathbb{Z}^{+}$ and $r$ as defined above, we have
 \begin{equation}
  \Prv{\mathcal{B}_n} \leq nr
 \end{equation}
\end{mylem}
\begin{IEEEproof}
 The proof proceeds by induction over $n$; the case $n=1$ holds by assumption. The induction step is as follows:
\begin{align}
 \Prv{\mathcal{B}_{n+1}} &= \Prv{\mathcal{B}_n \cup \mathcal{A}_{n+1}}\\
 &= \Prv{\mathcal{B}_n} + \Prv{\mathcal{A}_{n+1} \cap \mathcal{B}_n^c}\\
 &\leq \Prv{\mathcal{B}_n} + r \label{eq: comment1}\\
 &\leq (n+1)r. \label{eq: comment2}
\end{align}
where \eqref{eq: comment1} follows by assumption and \eqref{eq: comment2} uses the induction hypothesis.
\end{IEEEproof}

Now the statement of reliability follows easily. 

\begin{mythm}
\label{thm:reliability}
The error probability of the coding scheme satisfies $P_{\rm err}=O(L\,2^{-M^\beta}+L2^{-\frac12L^{\beta'}})$ for $\beta,\beta'>\frac12$. \end{mythm}

\begin{IEEEproof}
	For the polar source coding scheme, note that $\Prv{\mathcal{A}_i \cap \mathcal{B}_{i-1}^c} + x \in O(2^{-M^\beta})$, where $x$ is the probability that $\hat{V}_i \neq V_i$ given that a mistake previously occurred, but where we still give the correct $V^{i-1}$ to the decompressor. We can therefore upper bound $r$ in Lemma~\ref{lem: newerrLem} by $O(2^{-M^\beta})$~\cite{arikan10}.   Thus, the probability of incorrectly decoding any of the $|\Rset|$ $V_i$ is $O(L 2^{-M^\beta})$; this is $\eta'$ in Lemma~\ref{lem:wrongdecoder}. Lemma~\ref{lem:inputoutput}  and the properties of $\Rset$ give $\eta=O(L2^{-\frac12L^{\beta'}})$ for $\beta'>\frac12$, establishing the theorem.
\end{IEEEproof}

\section{Efficiency}
\label{sec:efficiency}
Here we consider the encoding, decoding, and construction complexity
of the coding scheme. Construction of the codes presented in
Section~\ref{sec:scheme} requires the random set $\Rset$ for the
shaper at the inner layer, and the deterministic sets (the $\Dset$)
for the tasks of compressing $V_i$ relative to side information
$Y^LV^{i-1}$ to determine the frozen bits at the outer layer. In
principle, these sets could be constructed by simulation, as
in~\cite{arikan09}. More satisfying would be a linear-time algorithm
along the lines of~\cite{talandvardy10, tal12} for the source coding
problem in which the variable to be compressed is not
uniformly-distributed. Presumably that algorithm can be adapted to the
problem of finding the frozen bits at the outer layer, as the
compressor actually used in Section~\ref{sec:scheme} is for an almost
uniformly-distributed random variable (cf.\
Lemma~\ref{lem:inputoutput}).  The complexity of constructing the
outer layer would then be $O(N)$, where $N=ML$.

\begin{myprop}\label{prop: encoding}
 The encoder has complexity $O\left(N \log N \right)$.
\end{myprop}
\begin{IEEEproof}
  The encoder consists of two parts, an outer and an inner
  encoder. The outer encoder consists of $|\Rset|$ multiplications
  with the matrix $G_M$, each requiring $O(M\log M)$
  operations~\cite{arikan09}.  Recalling the fact that
  $\left|\mathcal{R}_{\epsilon} \right|=O(L)$, we conclude that the
  complexity for the outer encoding is $O(M L \log M)$.

The inner encoder consists of $M$ rounds of the shaper $\S_{\left|\mathcal{R}_\epsilon \right|,L}$, for which the necessary multiplication with $G_L$ can be done in $O(L \log L)$. To construct $\hat{U}^L$, first note that by Definition~\ref{def: shaper} nothing has to be computed for $i\in\mathcal{R}_\epsilon$. For $i \not \in \mathcal{R}_{\epsilon}$, $Z_i$ can be generated using the likelihood ratio
\begin{equation} 
L^{(i)}\left(u^{i-1} \right) := \frac{\Prvcond{U_i=0}{U^{i-1}=u^{i-1}}}{\Prvcond{U_i=1}{U^{i-1}=u^{i-1}}},
\end{equation}
since $Z_i \sim \Bernoulli{L^{(i)}\left(u^{i-1}
  \right)/\left(L^{(i)}\left(u^{i-1} \right)+1\right)}$. All $L^{(i)}$
for $i\in[L]$ can be computed recursively with complexity $O\left(L
  \log L \right)$~\cite{arikan09}.  Thus, the inner encoding has
$O\left(M L \log L \right)$ complexity. Combining the inner and outer
encoding complexity establishes the claim.
\end{IEEEproof}

\begin{figure}[!htb]
	\centering
\def\gapx{1.7}
\def\gapy{.85}
\def\gaph{0.3}
\def\gaps{0.3}
\def\num{0.25}
\def\rad{0.2}
\def\radd{.23}
\def\spl{.2}

\begin{tikzpicture}[gr/.style={gray!75!black},x=\gapx cm,y=-\gapy cm]

\foreach \x in {1,2} {
  \foreach \y in {1,...,8} {
     }
}
\foreach \x in {1,2} {
  \foreach \y in {1,...,2} {
      }
}

\foreach \x in {1,...,8} {
   \draw (2,\x) -- (4.1,\x);
}

\foreach \x in {0,1} {
   \foreach \y in {0,4} {
   \fill (3.5-\spl+2*\x*\spl,\x+3+\y) circle (1.75pt);
   \draw (3.5-\spl+2*\x*\spl,\x+3+\y) -- (3.5-\spl+2*\x*\spl,\x+1-\radd+\y);
   \draw  (3.5-\spl+2*\x*\spl,\x+1+\y) circle [x radius=\rad cm, y radius =\rad cm];
   }   
   }
   
\foreach \x in {1,...,4} {
   \fill (2.5,2*\x) circle (1.75pt);
   \draw (2.5,2*\x) -- (2.5, 2*\x-1-\radd);
   \draw (2.5, 2*\x-1) circle [x radius=\rad cm, y radius =\rad cm];
}
   
\fill (1.25,8) circle (1.75pt);
\draw (1.25,8) -- (1.25,4-\radd);
\draw (1.25,4) circle [x radius=\rad cm, y radius =\rad cm];
\fill[gray] (1.25-.375,4) circle (1.25pt);
\fill[gray] (1.25-.375,8) circle (1.25pt);

\fill (0.1,5) circle (1.75pt);
\draw (0.1,5) -- (0.1,1-\radd);
   \draw (0.1,1) circle [x radius=\rad cm, y radius =\rad cm];
\fill[gray] (-.15,1) circle (1.25pt);
\fill[gray] (-.15,5) circle (1.25pt);


\draw (-.3,1) -- (2,1);
\draw (-.3,5) -- (2,5);
\draw (1-.3,4) -- (2,4);
\draw (1-.3,8) -- (2,8);

\foreach \x in {1,...,8} {
\node[anchor=west] at (4.1,\x) {$x_\x$};
}

\node[anchor=east] at (2,2) { $s^{(1)}_1$};
\node[anchor=east] at (2,3) { $s^{(1)}_2$};
\node[anchor=east] at (2,6) { $s^{(2)}_1$};
\node[anchor=east] at (2,7) { $s^{(2)}_2$};
\fill[gray] (1.75,1) circle (1.25pt);
\fill[gray] (1.75,4) circle (1.25pt);
\fill[gray] (1.75,5) circle (1.25pt);
\fill[gray] (1.75,8) circle (1.25pt);

\node[anchor=east] at (-.3,1) { ${t}^{(1)}_1$};
\node[anchor=east] at (-.3,5) { ${t}^{(1)}_2$};
\node[anchor=east] at (1-.3,4) { ${t}^{(2)}_1$};
\node[anchor=east] at (1-.3,8) { ${t}^{(2)}_2$};

\foreach \x in {1,...,8} {
\fill[gray] (2.9,\x) circle (1.25pt);
\fill[gray] (4,\x) circle (1.25pt);
}

\end{tikzpicture}
        \caption{\small Encoding circuit for the setup $L=4$, $M=2$, $K=2$, $\mathcal{E}_K = \left \lbrace 1,4 \right \rbrace$. Here $s_j^{(i)}$ denotes the $j$-th internally-generated bit of the shaper corresponding to the $i$-th super-channel, while $t_j^{(i)}$ is the $j$-th input to the $i$-th encoder at the outer layer. The small gray dots represent variables in the network and correspond to nodes in Fig.~\ref{fig: SC_decoder}.
}
        \label{fig: SC_encoder}
\end{figure}
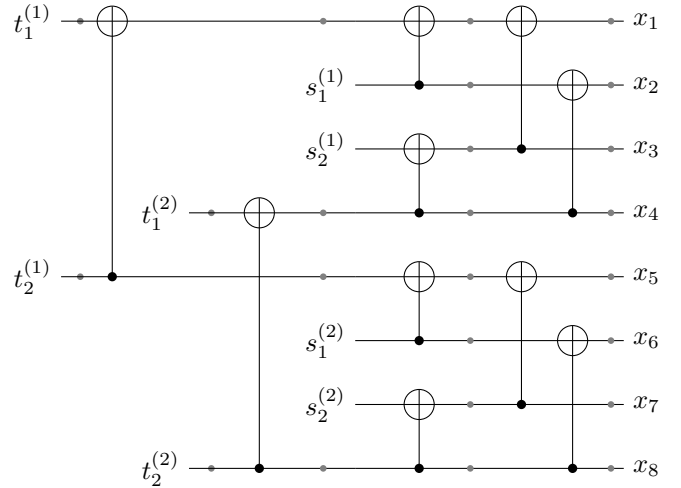

\begin{figure}[!htb]
	\centering
\def\gapx{1.7}
\def\gapy{1.2}
\def\gaph{0.3}
\def\gaps{0.3}
\def\num{0.25}
\def\la{0.05}

\begin{tikzpicture}[gr/.style={gray!75!black},x=\gapx cm,y=-\gapy cm]

\foreach \x in {1,2} {
  \foreach \y in {1,...,8} {
     \fill (\x+2,\y) circle (1.75pt);
     }
}
\foreach \x in {1,2} {
  \foreach \y in {1,...,2} {
     \fill (2*\x-2,4*\y-3) circle (1.75pt);
     \fill (\x,4*\y) circle (1.75pt);
      }
}

\foreach \x in {1,...,2} {
   \draw[] (3,\x) -- (4,\x+2) -- (3,\x+2) -- (4,\x) -- (3,\x);
   \draw[] (3,\x+4) -- (4,\x+2+4) -- (3,\x+2+4) -- (4,\x+4) -- (3,\x+4);

   }


\draw[] (0,1) -- (3,1);
\draw[] (2,1) -- (3,2);
\draw[] (0,1) -- (2,5);
\draw[] (2,5) -- (3,6);
\draw[] (0,5) -- (3,5);
\draw[] (2,1) -- (0,5);
\draw[] (1,4) -- (3,4);
\draw[] (2,4) -- (3,3);
\draw[] (1,4) -- (2,8);
\draw[] (2,8) -- (3,7);
\draw[] (1,8) -- (3,8);
\draw[] (1,8) -- (2,4);

\node[anchor=south] at (0,1) {1};
\node[anchor=south] at (2,1) {2};
\node[anchor=south] at (3,1) {3};
\node[anchor=south] at (4,1) {4};
\node[anchor=south] at (4+0.5*\la,3) {5};
\node[anchor=south] at (3,2) {6};
\node[anchor=south] at (4,2) {7};
\node[anchor=south] at (4,4) {8};
\node[anchor=south] at (2+0.5*\la,5) {9};
\node[anchor=south] at (3,5) {10};
\node[anchor=south] at (4,5) {11};
\node[anchor=south] at (4+\la,7) {12};
\node[anchor=south] at (3,6) {13};
\node[anchor=south] at (4,6) {14};
\node[anchor=south] at (4+\la,8) {15};
\node[anchor=south] at (5*\la,5) {16};
\node[anchor=south] at (1+3*\la,4) {17};
\node[anchor=south] at (2,4) {18};
\node[anchor=south] at (3-\la,3) {19};
\node[anchor=south] at (3-\la,4) {20};
\node[anchor=south] at (2-3.5*\la,8) {21};
\node[anchor=south] at (3-1.25*\la,7) {22};
\node[anchor=south] at (3-1.23*\la,8) {23};
\node[anchor=south] at (1+3.75*\la,8) {24};

\foreach \x in {1,...,8} {
\node[anchor=west] at (4,\x) {\small $y_\x$};
}

\node[anchor=north] at (3-3*\la,1) {\small $y_1,y_3$};
\node[anchor=north] at (3-3*\la,2) {\small $y_2,y_4$};
\node[anchor=north] at (2,1) {\small $y_1^4$};
\node[anchor=north] at (-\la,1) {\small $y_1^8$};
\node[anchor=north] at (2,5) {\small $y_5^8$};
\node[anchor=north] at (3-3.25*\la,5) {\small $y_5,y_7$};
\node[anchor=north] at (3-3*\la,6) {\small $y_6,y_8$};
\node[anchor=north] at (0,5) {\small $y_1^8,\hat{t}^{(1)}_1$};
\node[anchor=north] at (3,3) {\small $\blacktriangleup$};
\node[anchor=north] at (3+5*\la,4) {\small $y_2,y_4,s^{(1)}_1$};
\node[anchor=north] at (2+\la,4) {\small $\sqbullet$};
\node[anchor=north] at (1-5.75*\la,4) {\small $y_1^8,\hat{t}_1^{(1)2}$};
\node[anchor=north] at (1-5*\la,8) {\small $y_1^8,\hat{t}_1^{(1)2},\hat{t}_1^{(2)}$};
\node[anchor=north] at (3,7) {\small $\blackdiamond$};
\node[anchor=north] at (3+5*\la,8) {\small $y_6,y_8,s_1^{(2)}$};
\node[anchor=north] at (2,8) {\small $y_5^8,s_1^{(2)2},\hat{t}_1^{(1)2}$};


\node[] at (0,6.1) {\small $\blacktriangleup = y_1,y_3,\hat{t}^{(1)2}_1,s^{(1)}_1$};
\node[] at (-0.1,6.5) {\small $\sqbullet = y_1^4,s^{(1)2}_1,\hat{t}^{(1)2}_1$};
\node[] at (-0.05,6.9) {\small $\blackdiamond = y_5,y_7,s_1^{(2)},\hat{t}_2^{(1)}$};

%
%

\end{tikzpicture}
        \caption{\small Logical structure of the successive cancellation decoder for the setup $L=4$, $M=2$, $K=2$, $\mathcal{E}_K = \left \lbrace 1,4 \right \rbrace$ (compare with~\cite[Fig.\ 10]{arikan09}). Note that $\hat{t}_j^{(i)}$ denotes the $j$-th output of the $i$-th decompressor at the outer layer and $s_j^{(i)}$ denotes the $j$-th internal input to the $i$-th super-channel. The numbering of the nodes represents the order in which they get activated in the decoding process.}
        \label{fig: SC_decoder}
\end{figure}
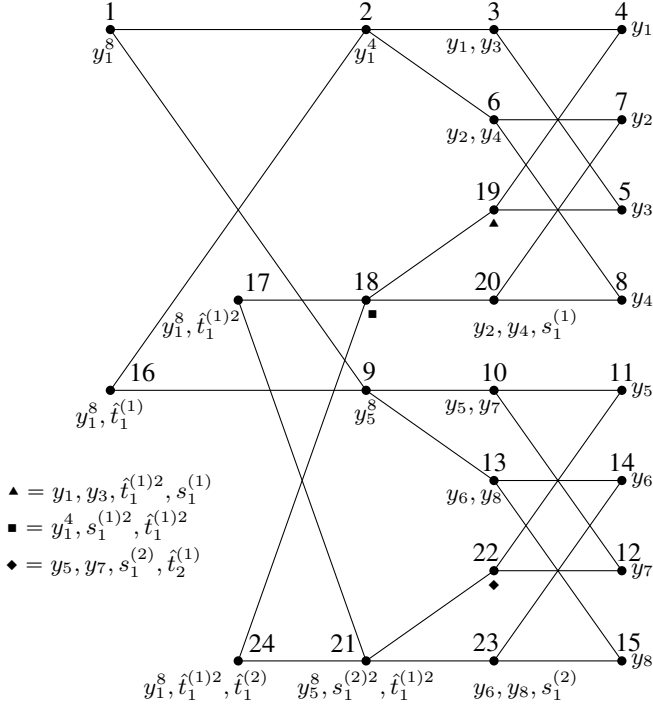

An important feature of the decoder is that the inner layer (super-channel) decompressors must be interleaved with the outer layer decompressors in order to ensure that all required variables are known at the appropriate steps. To illustrate, we explain in detail how the decoding is done for the setup $L=4$, $M=2$, $K=2$ and $\mathcal{E}_K=\left \lbrace 1,4 \right \rbrace$.\footnote{Recall that this implies that we have two compressors at the outer layer and two super-channels having a two bit input and a four bit output each. The second and third output of both shapers $\S_{2,4}$ are randomly distributed according to \eqref{eq:shaperdef} and are assumed to be known at the decoder.}  The logical structure of the successive cancellation decoder is shown in Figure~\ref{fig: SC_decoder}. Figure 10 of~\cite{arikan09} depicts a similar representation of the original successive cancellation decoder. To see the close affinity between the encoding and decoding process, Figure~\ref{fig: SC_encoder} visualizes the encoder for the setup defined above. 

Each node in Figure~\ref{fig: SC_decoder} is responsible for computing a LR arising during the algorithm; the parameters below each node represent the variables involved in the associated LR computation. Starting from the left we traverse the diagram to the right at whose border we can compute the LRs. Then we transmit the results back to the left. Here $\hat{t}_j^{(i)}$ denotes the $j$-th output of the $i$-th decompressor at the outer layer and $s_j^{(i)}$ denotes the $j$-th frozen input for the $i$-th super-channel.

The decoding begins by activating node $1$, which would like to compute the LR for $T_1^{(1)}$ given $Y_1^8$. For this it needs the LRs for the first inputs to the two super-channels, and so node $1$ activates node $2$, which is responsible for computing the LR for the first input to the first super-channel. This computation proceeds exactly as the usual successive cancellation decoder, recursively combining the LRs of the physical channels by calling node $3$ and then $6$. Assembling their results, node $2$ can compute its LR and transmits the result to nodes $1$ and $16$. Meanwhile, node $1$ has also requested the LR of node $9$, which performs the same calculation as node $2$ for the second super-channel, again forwards the result to nodes $1$ and $16$.  
Now node $1$ is able compute the final desired LR and can therefore guess $\hat{t}_1^{(1)}$. Having that value, node $16$ can guess $\hat{t}_2^{(1)}$, completing first decompressor of the outer layer. 

Node $16$ passes control to node $17$ in order to compute the LR for $T^{(2)}_1$.  This requires the LR for second inputs to the two super-channels, so nodes $18$ (and later $21$) are called. Node $18$ finishes the decompression of the first super-channel in the usual way, while node $21$ completes the decompression of the second super-channel. \emph{Neither of these can occur until the first outer layer decompressor is finished}. After the inner layer decompression is complete, node $17$ can guess $\hat{t}_1^{(2)}$ and node $24$ can finally guess $\hat{t}_2^{(2)}$, completing the second decompressor of the outer layer. 
In general, decompression of the $M$ different $k$-th inputs at the inner layer has to wait for the $(k-1)$-th decompressor to finish at the outer layer. 

\begin{myprop}
The decoder has complexity $O(N\log N)$.  
\end{myprop}

\begin{IEEEproof}
The decoder proceeds by employing, in sequence, the $|\Rset|$ decompressors for blocklength-$M$ compression of $V_i$ given $Y^LV^{i-1}$. This ensures that at all times the decoder has all the required previous inputs $V^{i-1}$. 
Each decompressor can be executed using $O(M\log M)$ operations, given the corresponding likelihood ratio (LR) of $V_i|Y^LV^{i-1}$. All such likelihoods can be computed in $O(L\log L)$ steps, and each of the $M$ super-channels requires its own likelihood calculation, as the values taken by $V^{i-1}$ can differ in each case. Using $|\Rset|=O(L)$, we find that the decompressor has complexity $O(N\log N)$.  
\end{IEEEproof}

\section{Derandomization}
\label{sec:derand}
Our coding scheme requires randomness at both the inner and outer layers. At the inner layer, the shaper randomly generates the inputs in $\Rset^c$, while the values of the frozen bits are to be chosen randomly at the outer layer. As the error probability of the coding scheme is the average over the possible assignments of these random values, at least one choice must be as good as the average, meaning a reliable, efficient, and deterministic coding scheme must exist. Thinking of the random choices as part of the code construction rather than the encoder, it follows by the Markov inequality that most choices will lead to coding schemes with these properties. 
Nonetheless, it is useful to consider derandomizing the construction, if only because  randomness can be difficult to generate. 

At the inner layer, the shaper of our coding scheme can be almost completely derandomized while incurring only a negligible overhead in error probability. Specifically, we alter the shaper so that for $i\in\Dset$, $\hat{U}_i$ is fixed to the most likely value of the distribution $U_i|U^{i-1}$, while the $Z_i$ corresponding to indices in the leftover set $\mathcal{A}_\epsilon:=\Rset^c\setminus\Dset$ are generated randomly as before. 
Since $|\mathcal{A}_\epsilon|=o(L)$, the required rate of randomness vanishes in the limit of large $L$. Nevertheless, the resulting scheme is still reliable; letting $P_{\rm err}'$ be the error probability of the coding scheme using the modified shaper and $P_{\rm err}$ as in Theorem~\ref{thm:reliability}, we have for $\beta<\frac{1}{2}$
\begin{mythm}
\label{thm:derandomize}
$P_{\rm err}'\leq P_{\rm err}\left(1+O\left(L\left(1-2^{-2^{-L^\beta}} \right)\right) \right).$
\end{mythm}
For the proof we need the following result
\begin{mylem} \label{lem: detrv}
 Let $R$ be a $\Bernoulli{p}$ distributed random variable with $p\in\left[\frac{1}{2},1 \right]$ such that $H(R)\leq \epsilon$. Then $p\geq 2^{-\epsilon}$.
\end{mylem}
\begin{IEEEproof}
Using $p\in\left[\frac{1}{2},1 \right]$ and some basic calculus we find
 \begin{equation}
  H(R)+\log \left(p\right)=  \left(1-p \right) \log \left(\frac{p}{1-p} \right) \geq 0.
 \end{equation}
Thus, by the premise,  $\epsilon \geq H(R) \geq -\log\left(p \right)$.
\end{IEEEproof}

\begin{IEEEproof}[Proof of Theorem~\ref{thm:derandomize}]
Let $\bar{u}^L$ denote the most likely sequence according to $P_{U^L}$. Then, by the union bound,
\begin{align}
P_{\rm err}&\geq P_{\rm err}'\,\,\Prv{{U}^L[\Dset]=\bar{u}^L[\Dset]}\\
&\geq P_{\rm err}' \big(1-\sum_{i\in\Dset}\Prv{U_i\neq \bar{u}_i}\big).
\end{align}
Each term in the summation may be written 
$\Prv{U_i\neq \bar{u}_i}=\sum_{u^{i-1}}\Prvcond{U_i\neq \bar{u}_i}{U^{i-1}=u^{i-1}}\Prv{U^{i-1}=u^{i-1}}$. But, from the fact that for $i\in\Dset$, $H(U_i|U^{i-1})\leq \epsilon$, according to Lemma~\ref{lem: detrv} the conditional probability is upper bounded by $1-2^{-\epsilon}$, regardless of the value of $U^{i-1}$. 
Using the size of $\Dset$ and form of $\epsilon$ completes the proof. 
\end{IEEEproof}

\section{Comparison with Gallager's Method} \label{sec: DiffGallager}
The main difference between the coding scheme presented in Section~\ref{sec:scheme} and Gallager's method \cite[p.208]{gallager68} is that the shaper $\S_{K,L}$ approximates the $L$-dimensional vector $X^L$ with $\hat X^L$, whereas Gallager's shaper $\S_G$ approximates the one-dimensional random variable $X$ through $\bar X$. Therefore, the super-channel $\W'_{K,L}$ consists of $L$ $\W$ channel uses, while $\W'_{G}$ consists of a single  $\W$ channel use. Note that $N$, as previously defined, denotes the number of physical channel uses. 
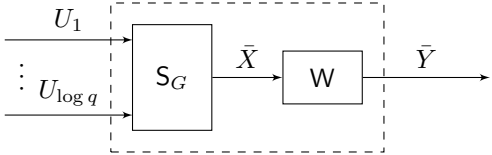
\begin{figure}[!htb]
	\centering

	
\tikzstyle{blocks} = [draw, rectangle, 
    minimum height=4em, minimum width=3em, node distance=2.5cm,anchor=center]
\tikzstyle{blockw} = [draw, rectangle, 
    minimum height=2em, minimum width=3em, node distance=2.5cm,anchor=center]
\def\gap{1}

\begin{tikzpicture}[auto,>=latex']
	
    \node (input) at (-3.35*\gap,0) {};
    \node [blocks] (shaper) at (-1*\gap,0) {$\S_{G}$};
    \node [blockw] (channel) at (1*\gap,0) {$\W$};
    \node (output) at (3.35*\gap,0) {};

    \node (oneb) at (-3.35*\gap,0.5) {};
    \node (onef) at (-1.4*\gap,0.5) {};
    \node (twob) at (-3.35*\gap,-0.5) {};
    \node (twof) at (-1.4*\gap,-0.5) {};

    \draw [->] (oneb) -- node {{$U_1$}} (onef); 
    \draw [->] (twob) -- node {$U_{\log q}$} (twof);
    \draw [->] (shaper) -- node {$\bar X$} (channel);
    \draw [->] (channel) -- node {$\bar Y$} (output);
   
    \node [rotate=90] at (-3*\gap,0.04) {$\dots$};

    \node[draw,dashed,fit=(shaper) (channel),inner sep=8pt] (super) {};

\end{tikzpicture}
\caption{\small Gallager's super-channel $\W'_{G}:=\W \circ \S_G$. A
  $q$-ary input $U^{\log q}$ (with $q=2^m$ for $m\in \mathbb{Z}^{+}$)
  whose elements are i.i.d. $\Bernoulli{\frac12}$ distributed is
  shaped into a rational approximation to $X$, i.e. $\bar X \sim
  \Bernoulli{k/q}$ where $k\in \mathbb{Z}^{+}$ and $k/q\approx p$.}
        \label{fig: Gallager}
\end{figure}

Gallager's method is based on the approximation of $p$ by $k/q$, where $k\in \mathbb{Z}^{+}$ and $q=2^m$ for $m\in \mathbb{Z}^{+}$. 
For a binary channel whose optimal input is $\Bernoulli{p}$ for an irrational $p$ requires, in principle, an infinitely-large $q$. The crucial question is how fast $q$ must increase relative to $N$. 

It is simple to verify that   
\begin{equation}
 \delta\left(X,\bar X \right) = \min \limits_{k \in \mathbb{Z}^{+}} \left|p-\frac{k}{q} \right| \leq \frac{1}{2q}. \label{eq: approxGallager}
\end{equation}
Then the polar coding scheme introduced in \cite{abbe_polar_2011}\footnote{Note that in terms of complexity this scheme improves the scheme initially proposed for Gallager's method~\cite{sasoglu2009}.} can be applied to the super-channel $\W'_{G}$; it has an encoding complexity of $O\left(\log q \cdot N\log N \right)$ and a decoding complexity $O\left(q\log q\cdot N\log N\right)$. Furthermore the probability of error behaves as $O\left(\log q\cdot 2^{-N^\beta} \right)$ for $\beta <\frac{1}{2}$. 
Using this scheme leads to  
\begin{myprop}
Gallager's scheme achieves a rate of  $C(\W)-O(\frac{1}{q} \, \log q )$ for channels $\W$ with an irrational optimal input distribution. 
\end{myprop}
\begin{IEEEproof}
Using \eqref{eq: approxGallager} and the monotonicity of the variational distance gives 
\begin{equation}
 \delta\left((X,Y ), (\bar X,\bar Y ) \right)\leq \frac{1}{2q}. \label{eq: tworv}
\end{equation}
 From \cite[Lemma 2.7]{csiszarkorner81}, \eqref{eq: approxGallager} and the monotonicity of the variational distance we obtain $\left|\Hh{X}- \Hh{\bar X} \right| \leq \frac{1}{q} \log \left(2q \right)$ and $\left|\Hh{Y}- \Hh{\bar Y} \right| \leq \frac{1}{q} \log \left(2q \right)$. The same reasoning applied to \eqref{eq: tworv} gives $\left|\Hh{X,Y}-\Hh{\bar X, \bar Y} \right| \leq \frac{1}{q} \log \left(4q \right)$. Using the chain rule leads to $\left|\Hcond{Y}{X}-\Hcond{\bar Y}{\bar X}\right|\leq \frac{2}{q} \log q + \frac{3}{q}$.
Thus,
\begin{align}
 &\left|I(X:Y)-I(\bar X:\bar Y) \right|\nonumber\\ 
  &\hspace{10mm}= \left|\Hh{Y}-\Hcond{Y}{X}-\Hh{\bar Y}+\Hcond{\bar Y}{\bar X} \right|\\
  &\hspace{10mm}\leq \frac{3}{q} \log q + \frac{4}{q} = O\left(\frac{1}{q} \log q \right).\qedhere
\end{align}
\end{IEEEproof}

Table~\ref{table: GallagervsOurs} summarizes the differences between Gallager's method and the new scheme. What can be said is that the new method has better complexity but generally worse error probability than Gallager's method. If $q$ is chosen to increase slowly (e.g. $q=O\left( \log N \right)$), Gallager's scheme works with a comparable complexity and superior error probability, but the rate converges much more slowly to the capacity. Choosing $q$ to increase quickly (e.g. $q=O\left(N \right))$, on the other hand, the rates of both schemes converge comparably fast to the capacity, but the reduced error rate of the Gallager scheme is offset by the essentially quadratic complexity. 
\begin{table}[!t]
\renewcommand{\arraystretch}{1.3}
\caption{\small Summary of the important parameters for the two different schemes for $M=L=\sqrt{N}$. Recall that $\beta < \frac{1}{2}$.}
\label{table: GallagervsOurs}
\centering
\begin{tabular}{c c c}
\hline
& Gallager's scheme & Our scheme \\
\hline
Rate & $C-O\left(\frac{1}{q} \log q \right)$ & $C-\frac{o(N)}{N}$\\
Complexity & $O\left(q \log q\cdot  N \log N \right)$ & $O\left(N \log N \right)$\\
Error probability & $O\left(\log q\cdot  2^{-N^\beta} \right)$ & $O\Big(\sqrt{N}2^{-\frac{1}{2} N^{\frac{\beta}{2}}}\Big)$\\
\hline
\end{tabular}
\end{table}

\section{Discussion} \label{sec: discussion} We have used the
polarization phenomenon to construct a distribution shaper and shown
how it can be concatenated with a version of polar channel codes to
yield a coding scheme which achieves the capacity of any binary-input
DMC. For DMCs with arbitrary input sizes, we can again employ
multilevel coding.

\subsection{Possible Modifications}
Several modifications to our coding scheme are possible. In principle, neither layer need be based on polar codes, and other randomness extractors and coding schemes which are in some way advantageous could equally-well be used. For instance the ``invertible extractors'' of~\cite{cheraghchi_invertible_2012} may prove suitable (provided such invertible extractors can be used for shaping).  However, designing outer layer codes and decoding them efficiently may prove challenging, as the properties of the super-channel may be difficult to determine.  One simple modification to the outer layer, concatenation with Reed-Solomon codes, can lead to an improved error rate at the outer layer with almost no cost in computational complexity~\cite{baskshi10}.

Within the realm of polar codes, one could use q-ary codes for the
outer layer~\cite{sasoglu2009,abbe_polar_2011}, instead of multilevel
coding. Similarly, q-ary polar source coding could be used to design
shapers for channels with non-binary input~\cite{karzand10}. Following
the analysis of Section~\ref{sec: DiffGallager}, it can be verified
that using a $2^K$-ary polar code at the outer layer leads to a worse
complexity ($O\left(2^L L M \log M \right)$ as opposed to $O\left(L M
  \log M \right)$), while the error probability remains the same
(namely $O\left(L 2^{-M^\beta} \right)$ for $\beta <\frac{1}{2}$).

At the outer layer, $O(LM)$ bits of randomness are nominally needed to determine the frozen inputs. However, as the capacity of the super-channel is presumably achieved by a uniform input (or non-uniform inputs add only $o(L)$ terms to the mutual information), perhaps it is possible to show that it is  indeed a symmetric channel (or at least approximately so), so that all choices of frozen bits are equivalent, enabling a deterministic choice~\cite[Section VI]{arikan09}.

\subsection{Applications}

It would be interesting to adapt the method presented here to other
settings. In the realm of binary discrete memoryless channels, the
shaping gap---the penalty in lost capacity for working with a uniform
input distribution instead of the optimal one---never exceeds
6\%~\cite{shulman_uniform_2004}, so our method is of limited practical
utility for binary channels. However, the shaping gap can be
arbitrarily large in other scenarios, e.g.\ input letters of differing
duration~\cite{jimbo79}, channels with power constraints on the input
symbols \cite{blahut72}, and multi-user channels with cross-talk
\cite{ratzer03}.

One possible application for the new scheme is the $m$-user MAC, where the new method might be used to achieve rate regions with non-uniform inputs~\cite{sasoglu09_2}, \cite{abbetelatar10}. Our method should also be applicable to the construction of quantum polar codes~\cite{wilde_polar_2011,renes_efficient_2011}. Perhaps most interesting is the benefit our scheme brings to the AWGN channel with an average power constraint, which we discuss in more detail in the remainder of this section. 

The capacity of the AWGN channel, with inputs constrained to a finite average power, can in principle be achieved by discretizing the inputs and employing codes for DMCs. Polar codes offer an efficient, capacity-achieving scheme, as described in~\cite{abbe_polar_2011}.  Our coding scheme improves on that method. Let $\nu \geq 0$ and $Z \sim \mathcal{N}(0,\nu)$, we define for $m\in \mathbb{Z}^{+}$,
\begin{align}
 C_{m,1}&:= \sup \limits_{\mathsf{E}\left[X^2 \right]\leq 1, \, \,
                          \left|\supp\left(P_X\right)\right|\leq 2^m} I(X:X+Z) \label{eq: ourapprox} \\ 
 C_{m,2}&:= \sup \limits_{\mathsf{E}\left[X^2 \right]\leq 1, \, \,
                          X \textnormal{ is }m\textnormal{-dyadic}} I(X:X+Z). \label{eq: abbesapprox}
\end{align}
These are the respective capacities for coding with power-constrained, but otherwise arbitrary constellations of $2^m$ discrete points or power-constrained constellations described by an $m$-dyadic discrete random variable $X$, whose probability distribution has the form $P_X(x)=k\,2^{-m}$ for $k\in \mathbb{Z}^{+}$ and $x\in \supp(P_X)$. In the limit of large $m$, both quantities approach the true capacity of the AWGN channel,    $C:=\frac{1}{2}\log\left(1+\SNR \right)$, whose optimal input distribution is simply $X\sim \mathcal{N}\left(0,1 \right)$. 

The convergence rate of $C_{m,2}$ is exponential in $m$,
\begin{align}
 C-C_{m,2} \leq \SNR \,2^{-m}, \label{eq: abberes}
\end{align}
and this rate is shown to be achievable with polar codes in~\cite{abbe_polar_2011}. 
Using our new coding scheme we can relax the constraint of $X$ being $m$-dyadic to $\left|\supp\left(P_X\right)\right|\leq 2^m$ and thus we can achieve $C_{1,m}$ using codes with the same complexity. Indeed, the benefit of the improved approximation $C_{m,1}$ can be large: 
According to \cite[Theorem 8]{wu10}, using a Gauss quadrature constellation leads to double exponential convergence rate,
\begin{equation}
C-C_{m,1} \leq 4 \left(1+\SNR \right) \left(\frac{\SNR}{1+\SNR} \right)^{2^{m+1}}. \label{eq: gaussquad} 
\end{equation}

\sloppy
\bibliographystyle{ieeetr}           
\bibliography{header,bibliofileLinksNew}    

\end{document}